\documentclass[11pt,a4paper,onecolumn]{article}
\usepackage[top=2.54cm, bottom=2.54cm, left=2.54cm, right=2.54cm, headsep=0.15cm]{geometry}
\usepackage[utf8]{inputenc}
\usepackage{whitepaper}

\DeclareSourcemap{
  \maps[datatype=bibtex]{
    \map{
      \step[fieldset=urldate, null]
    }
  }
}

\DeclareSourcemap{
  \maps[datatype=bibtex]{
    \map{
      \step[fieldset=editor, null]
    }
  }
}

\title{Extragalactic planetary nebulae -- tracers of kinematics and stellar populations out to 100~Mpc}
\date{December 15, 2025}
\author[1,2,3]{Johanna Hartke}
\author[4]{Magda Arnaboldi}
\author[5]{Claudia Pulsoni}
\author[6]{Souradeep Bhattacharya} 
\author[7]{Martin Bureau}
\author[8]{Enrico Congiu}
\author[7]{Guy Flint} 
\author[5]{Ortwin Gerhard} 
\author[9]{Martin Roth} 
\author[9,10]{Azlizan Soemitro}
\author[7,4]{Chiara Spiniello}
\author[11]{Lucas Valenzuela}
\author[9]{Peter Weilbacher} 
\author[7]{Nancy Yang}

\affil[1]{Finnish Centre for Astronomy with ESO, (FINCA), University of Turku, Finland}
\affil[2]{Tuorla Observatory, Department of Physics and Astronomy, University of Turku, Finland}
\affil[3]{Turku Collegium for Science, Medicine and Technology (TCSMT), University of Turku, Finland}
\affil[4]{European Southern Observatory, Garching, Germany}
\affil[5]{Max-Planck-Institut f\"{u}r Extraterrestrische Physik, Garching, Germany}
\affil[6]{Centre for Astrophysics Research, Department of Physics, Astronomy and Mathematics, University of Hertfordshire, UK}
\affil[7]{Department of Physics, University of Oxford, UK}
\affil[8]{European Southern Observatory, Vitacura, Chile}
\affil[9]{Leibniz-Institut für Astrophysik Potsdam (AIP), Germany}
\affil[10]{Institut für Physik und Astronomie, Universität Potsdam, Germany}
\affil[11]{Universitäts-Sternwarte, Fakultät für Physik, Ludwig-Maximilians-Universität München, Germany}

\begin{document}
\maketitle

\newpage
\textbf{Abstract:} 
    Extragalactic planetary nebulae (xPNe) in galaxies beyond the Local Universe serve as discrete tracers for studying the element abundances and kinematics of galaxies covering a wide range of morphologies and masses at a variety of angular distances, from the centre well out into their haloes. They are direct stellar probes to identify the galaxy progenitors of haloes and the intracluster light. Even with new facilities, reaching higher angular resolution and sensitivity, xPNe are the only stellar tracers that can be directly and singularly detected and characterised out to 100 Mpc distance, making them crucial for tracing halo and intracluster light assembly. New wide-field spectroscopic instruments at 10+meter-class telescopes will allow the unprecedented characterisation of xPN populations from galaxy centres to their diffuse outskirts. Panoramic integral-field spectroscopy will enable the simultaneous study of xPN and stellar population properties, establishing their use as age- and metallicity-tracers while also improving post-AGB stellar evolution models.

\section*{Context}
According to cosmological models of structure formation, galaxies and clusters grow hierarchically through the mergers and accretion of smaller subsystems \autocite{1984Natur.311..517B}. In this paradigm, accreted stars from minor mergers will be preferentially deposited in the outskirts of galaxies. In the outer haloes of galaxies, where dynamical timescales are long, and in the region between galaxies filled with intra-cluster or intragroup light (IC/IGL), the fossil record of this assembly process is preserved in the spatial distribution, chemical abundances, and kinematics of individual stars \autocite{2013MNRAS.434.3348C, 2016ApJ...833..158C}.

\begin{wrapfigure}{r}{0.33\textwidth}
  \centering
  \vspace{-\baselineskip}
    \includegraphics[width=0.32\textwidth]{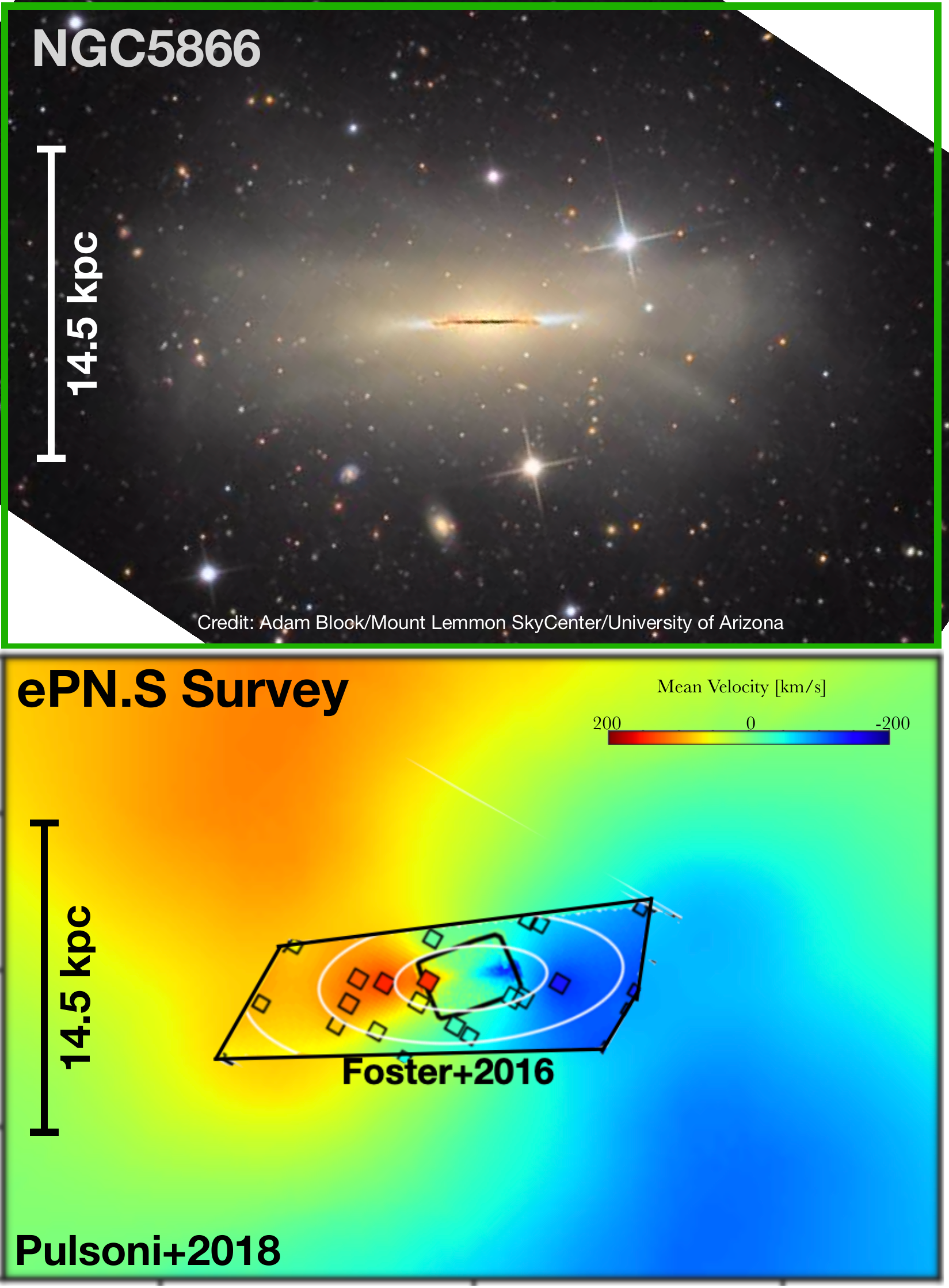}
  \vspace{-5pt}
  \caption{Large-scale velocity field of NGC~5866 derived from PNe in the scope of the ePN.S survey \autocite{PNS_Pulsoni} compared to the central stellar kinematics from the ATLAS3D\autocite{2011mnras.413..813c} and the SLUGGS\autocite{2016mnras.457..147f} surveys. Figure adapted from \cite{PNS_Pulsoni, 2016mnras.457..147f}.}
  \vspace{-15pt}
  \label{fig:Pulsoni}
\end{wrapfigure}

It has become clear in recent years that important tests of cosmological models are possible with kinematic observations targeting the diffuse outer regions of galaxy haloes and individual low surface brightness (LSB) galaxies \autocite{2018ApJ...856L..31T}. In practice, this information is often hidden at surface brightnesses values below the sky \autocite{2017ApJ...834...16M}, making the kinematic study of haloes and the ICL with integral-field spectroscopy (IFS) very challenging.
This challenge can be overcome by using planetary nebulae (PNe) as discrete tracers. 
PNe occur in the late stages of low-mass stellar evolution, and their typical spectral characteristics\footnote{Absence of an optical continuum and strong emission lines, with up to $15\%$ of the radiation of the central star predicted to be re-emitted in a single optical line $\lambda5007$ Å [\ion{O}{iii}] \autocite{1992ApJ...389...27D}.} allow for their detection in galaxies as distant as 100~Mpc \autocite{2005ApJ...621L..93G}.  

In deep [\ion{O}{iii}] images, PNe are readily detectable in extended haloes and the transition regions between individual galaxies and the ICL, both spatially and kinematically. They are sometimes observed to follow extended LSB structures, such as tidal streams, plumes and shells \autocite{2015A&A...579L...3L, fensch_2019A&A...625A..77F, 2021A&A...647A.130B, 2023MNRAS.522.6010B}. Once identified, they are ideal targets for multi-object spectrographs on 10+m-class telescopes \autocite{2015A&A...574A.109W, 2018A&A...620A.111L}. In addition to the information provided by observations of their specific frequencies and spatial distribution, PNe are unique tools for measuring the angular momenta and orbital motions of stars in the outermost regions of galaxies, where the diffuse halo is beyond the reach of absorption line spectroscopy \autocite{PNS_Pulsoni}, as illustrated in Fig.~\ref{fig:Pulsoni}.

Besides their extended use as kinematic tracers, the study of the PN populations in galaxies of all types links the late phases of the stellar evolution with the star formation histories, ages and metallicities of their parent stellar populations. Both the PN specific frequency ($\alpha$-parameter, linked to the PN visibility time scale) and the morphology of the PN luminosity function, including its bright cut-off and gradients, are avenues towards a better physical understanding of these enigmatic end products of stellar evolution and mass loss. 
PNe are furthermore ubiquitous tracers of distances and element abundances (see companion white paper by Arnaboldi et al.) in the outer LSB regions of galaxies which can be studied with 10+ meter facilities, equipped with new wide-field multi-fibre and integral field spectrographs. In the following sections, we will provide a short summary of the science cases that can be targeted using extragalactic PNe (xPNe) as example of resolved stellar populations with the next facilities.

\section*{xPNe as discrete tracers of halo and IGL/ICL assembly}
Unlike globular clusters (GCs), PNe are drawn from the main stellar population of galaxies with older stellar populations like early-type galaxies (ETGs). Hence, their number density and kinematics directly trace those of the stars in the outer haloes of galaxies and in the ICL/IGL, in the regime of extremely low surface brightness. 
There, PNe are invaluable dynamical probes reaching well into the dark matter-dominated regions \autocite{delorenzi_2008MNRAS.385.1729D, napolitano_2009, 2009A&A...502..771D, 2018A&A...620A.111L, 2022A&A...663A..12H}. 
Current facilities have also enabled the use of xPNe as tracers of the halo orbital structure of ETGs, unveiling triaxial-spheroidal shapes \autocite{PNS_Pulsoni, 2021A&A...647A..95P} and orbital anisotropies \autocite{2003Sci...301.1696R, 2009MNRAS.395...76D}. They robustly quantified the angular momentum content of ETGs, establishing important constraints in the evolution of these galaxies and the role of mergers \autocite{PNS_Pulsoni, pulsoni_2023A&A...674A..96P}. 

The $\alpha$-parameter (the number of PNe per unit bolometric luminosity), incarnates a first link between PNe and the underlying stellar population properties as it has been found to depend on the metallicity and star formation history. In particular, studies investigating the outermost haloes of ETGs have observed increasing $\alpha$-parameter values with the distances from the centres caused by the increased contribution of the ICL/IGL and its blue, metal-poor stars \autocite{2009A&A...502..771D, 2013A&A...558A..42L, 2018A&A...616A.123H, 2020A&A...642A..46H}.

However, these detailed studies have been limited to a few nearby groups and clusters ($\leq 25$ Mpc). What is missing is a systematic survey of xPNe in group and cluster environments to provide a kinematic counterpart of the ongoing deep imaging efforts with facilities such as Rubin LSST and the Euclid mission, unveiling unprecedentedly detailed LSB structures surrounding galaxies \autocite{2025A&A...697A..13K}. PNe are emission-line objects, and thus amenable to study at moderately high spectral resolutions ($R\gtrsim5000$) and can yield radial velocities accurate to $\sim5$ km/s, ideal for studying the orbital properties and dark matter content of dynamically ``cold'' stellar systems, like dwarf galaxies and stellar streams. 
Future wide-field ($\mathcal{O}(\mathrm{deg}^2)$) spectroscopic facilities will be a game-changer to identify the progenitors of the haloes and ICL/IGL via the spectroscopy of PNe. The resulting spectra will not only provide important kinematic information, but can be used to derive stellar abundances, expansion velocities, and even central star masses \autocite{2008ApJ...674L..17A}. 

\section*{Linking xPNe with their underlying stellar populations}

It is not straightforward to link xPN properties with that of the underlying stellar populations. The first reason for this is classical PN surveys using narrow-band imaging or slitless spectroscopy being ``blind'' in the central regions of galaxies. There, even the strongest [\ion{O}{iii}] emission from PNe is outshone by the galaxy. But this is also the location where stellar population properties can be most readily determined via absorption spectroscopy, limiting systematic studies to correlate e.g. $\alpha$-parameters to the metallicities of the underlying stellar populations.
Integral field spectrographs have overcome this limitation, as they allow the pixel-by-pixel decomposition of the bright central regions of galaxies into the respective stellar and nebular contributions, finally facilitating the detection of xPNe even in the centres of galaxies \autocite{roth_2004ApJ...603..531R, 2011MNRAS.415.2832S} and to measure accurate stellar ages and metallicities from the \textit{same data} \autocite{Cappellari04, 2005MNRAS.358..363C, 2018ApJ...854..139C}. 

The second reason concerns the theoretical understanding of post-AGB stellar evolution: until very recently, post-AGB models were at odds with the large quantity of bright PNe in ETGs \autocite{2004A&A...423..995M, 2012Ap&SS.341..151C}. Using faster-evolving model tracks \autocite{2016A&A...588A..25M} and more detailed prescriptions for modelling PNe in cosmological simulations from simple stellar populations of different masses, ages, and metallicities, it is now possible to simulate bright PNe in old and metal-rich stellar populations \autocite{2025A&A...699A.371V}, akin to those of massive ETGs, directly confronting the observations \autocite{2025arXiv250910175S}. However, several open questions remain concerning the role of different model ingredients, such as the initial-to-final mass relation, and dust and circumnebular extinction. 

Enabling large integral-field spectroscopic surveys for xPNe and stellar population properties is extremely timely. Establishing correlations between the two tests predictions from state-of-the-art simulations of PNe in cosmological simulations and improves our understanding of post-AGB stellar evolution. Furthermore, they can be used to elevate xPNe in LSB haloes and the ICL/IGL, where ages and metallicities will remain largely inaccessible, to indirect stellar population property tracers. 

Current instruments such as MUSE at the \textit{Very Large Telescope}, with its broad wavelength coverage and high spatial resolution, have already opened up a tremendous discovery space for xPNe, especially in late-type galaxies \autocite{2017ApJ...834..174K, 2022MNRAS.511.6087S, congiu_2025A&A...700A.125C}, which were historically under-represented due to the presence of many other [\ion{O}{iii}] emitters such as \ion{H}{ii} regions and supernova remnants, which makes the identification of PNe from narrow-band surveys less efficient. Different techniques have been established for the robust detection of PNe from integral-field spectroscopic data \autocite{spriggs_2020A&A...637A..62S, 2021ApJ...916...21R, 2025MNRAS.tmp.1932E} that can be applied to observations of galaxies of all types. Still, to date, studies are limited by the small ($\mathcal{O}(1\,\mathrm{arcmin}^{2})$) field-of-view of current and upcoming (e.g. BlueMUSE \autocite{2019arXiv190601657R}) state-of-the-art integral field spectrographs. Wide-field ($\mathcal{O}(10\,\mathrm{arcmin}^2)$) integral-field spectrographs on 10+meter-class telescopes will be transformative for the \textit{simultaneous} study of xPNe and their host stellar populations in galaxies of all types well beyond 25~Mpc.

\section*{Opportunities for xPNe surveys with new facilities}
Systematic surveys for xPNe from the centres of galaxies to the surrounding diffuse light will be transformative for our understanding of halo and IGL/ICL assembly as traced by xPNe, and also for establishing a relation between xPNe and their host stellar populations. Spectroscopic surveys for xPNe will be complemented by deep pencil-beam studies with instruments such as MICADO \autocite{2024SPIE13096E..11S} and HARMONI \autocite{2016SPIE.9908E..1XT}  at the \textit{Extremely Large Telescope}. First results from wide-field imaging with facilities such as Rubin LSST and the Euclid mission are extremely promising for synergies on tracing LSB accreted structures, with further missions with the goal of exploring the nearby LSB universe such as ARRAKIHS \autocite{2024SPIE13092E..0QC} already on the horizon. 

Currently, no ground- (nor space-) based telescope and instrument combination exists with which the above science goals for the study of xPNe can be fully realised. A new ground-based facility combining a large collecting area (10+meter-class telescope), with panoramic integral-field units, highly multiplexed multi-object spectrographs with moderate spectral resolution ($R\sim5000$), and ideally also a ground-layer adaptive optics system to boost the detection and signal-to-noise ratio of point-like sources \autocite{2024ApJS..271...40J}, is the only way to advance our understanding of galaxy assembly with xPNe.
\vspace{3pt}
\\
\textbf{References:}
\printbibliography[heading=none]

\end{document}